\documentclass[12pt]{article}

\usepackage{amssymb,amsmath}

\DeclareMathOperator{\Si}{Si}
\DeclareMathOperator{\Ci}{Ci}

\begin{document}

\begin{titlepage}

\begin{flushright}
arXiv:2007.00133
\end{flushright}
\vskip 2.5cm

\begin{center}
{\Large \bf Problems with Lorentz Violation Originating From\\
A Cosmologically Varying Pseudoscalar Field}
\end{center}

\vspace{1ex}

\begin{center}
{\large Sapan Karki and Brett Altschul\footnote{{\tt altschul@mailbox.sc.edu}}}

\vspace{5mm}
{\sl Department of Physics and Astronomy} \\
{\sl University of South Carolina} \\
{\sl Columbia, SC 29208} \\
\end{center}

\vspace{2.5ex}

\medskip

\centerline {\bf Abstract}

\bigskip

Lorentz- and CPT-violating models of electrodynamics with Chern-Simons terms are typically plagued by
various sorts of instabilities. However, when the Chern-Simons term arises from a slow time variation in
a pseudoscalar field with an axion-like electromagnetic coupling, the
total energy of the theory is bounded below. We examine
the behavior of such a theory, finding that in a systematic power series expansion of the magnetic and pseudoscalar fields,
singularities appear in the field profiles. Some of the questionable behavior can be cured by taking a fully
nonperturbative approach, but other problematic terms remain. This may be an indication that Cerenkov-like radiation
will automatically carry away energy from a moving charge, preventing a charge from moving with uniform
velocity over extended distances.

\bigskip

\end{titlepage}

\newpage

\section{Introduction}

One of the key themes of modern fundamental physics is symmetry. The importance of this
topic extends to situations with both exact and also, interestingly, broken symmetries. In fact,
many operations that may have initially appeared to be exact symmetries of
elementary particle physics have since proved to represent merely approximate symmetries.
These approximate symmetries (such as isospin or parity) and how they were actually violated
provided crucial insights into the structure of the standard model at successively deeper levels.

In order to bring the standard model describing particle physics and the general theory of
relativity together, some new physics beyond what we currently understand must exist. Whatever
new physics exists at more fundamental scales that we have not yet probed might also involve further
new forms of symmetry breaking. It is interesting to question whether new fundamental physics
might yet break some of the seemingly strongest symmetries that we have thus far encountered---such
as Lorentz and CPT symmetries. Lorentz and CPT invariance are related to some quite
basic properties of field theories: spatial isotropy, Lorentz boost invariance, and
unitary time evolution. Both Lorentz and CPT symmetries are basic building
blocks of the standard model and of general relativity, and they are tied together by the
CPT Theorem, which in its most general form requires CPT invariance in a Lorentz-invariant,
stable, unitary quantum field theory~\cite{ref-greenberg}.
However, there is no guarantee that they
should continue to hold exactly in a more fundamental theory; and, in fact, a number of the
schematic frameworks that have been proposed to deal with quantum gravity suggest that they
may support Lorentz or CPT symmetry breaking.

Experimentally, there has thus far been no convincing evidence for Lorentz or CPT violation. If
violations of one or both of these symmetries are ever uncovered, that would be a discovery of
extraordinary significance. However, even in the absence of physical violations of these symmetries,
theories with Lorentz violation, CPT violation, or other similarly exotic features can be
extremely informative for our understanding of how the kinds of field theories that we use
to explain the universe's fundamental interactions may behave in general. Such unusual theories
may provide unexpected insights about the general behavior of the field theoretic framework.

For dealing with questions about possible Lorentz and CPT violation in the interactions of standard model
particles, the most natural formalism is effective quantum field theory.
A general effective field theory that entails all possible Lorentz- and CPT-violating additions that may be made
to the standard model without introducing any additional conjectural quanta has been described.
This theory, known as the standard model extension (SME), contains operators formed out of
the usual standard model fields, but without the usual requirement that the action be a Lorentz
scalar~\cite{ref-kost1,ref-kost2}. The minimal SME, containing the finite number of Hermitian, local,
gauge-invariant, and
renormalizable~\cite{ref-kost3,ref-berr,ref-collad-3,ref-collad-2,ref-collad-1,ref-gomes,ref-anber,ref-ferrero3}
operators that can be formed in this way, offers
an extremely useful test theory for parameterizing the results of experimental Lorentz and CPT tests.

The SME, as an effective low-energy theory, can be used to describe the experimentally accessible
limits of a more fundamental theory. The SME itself can accommodate Lorentz and CPT violation in
low-energy observables, regardless of how the symmetries are broken in the more fundamental underlying
theory. One way in which SME operators could naturally be generated is by cosmological evolution.
The universe has a naturally preferred reference frame, in which the cosmic microwave background is at rest.
If the fundamental dimensionless constants of the standard model (such as the fine structure
constant $\alpha$, or the ratio of the electron mass to the quantum chromodynamics scale, $m_{e}/\Lambda_{{\rm QCD}}$)
are varying with time, due to some slow-acting dynamics, there must also be attendant Lorentz
violation; if $\partial_{\mu}\alpha\neq0$, then then $\partial_{\mu}\alpha$ gives a preferred spacetime direction.
In fact, varying $\alpha$ can quite naturally give rise, via radiative corrections, to SME operators
describing photon-sector Lorentz violation~\cite{ref-ferrero1}.

Another type of SME Lorentz violation that could be generated in a spacetime with slowly varying
cosmological solutions is an electromagnetic Chern-Simons term~\cite{ref-kost20}. The behavior of the
kind of Chern-Simons term that might be generated by a varying coupling in this way will be the principal 
subject of this paper.
In the usual SME Lagrange density, the Chern-Simons
term is rather peculiar, since it depends on the vector potential $A^{\mu}$, rather than just on the
electromagnetic field strength $F^{\mu\nu}$---and in such a way that the term is not quite gauge invariant.
The structural subtleties associated with this kind of term made it a source of significant controversy, in particular
in regard to whether there could be a radiatively generated Chern-Simons term in a Lorentz- and CPT-violating
quantum field theory. It was found that different regulators
applied to superficially divergent loop integrals could lead to different finite
radiative corrections to the
Chern-Simons term~\cite{ref-coleman,ref-jackiw1,ref-victoria1,ref-chung1,ref-chung2,ref-chen,ref-chung3,
ref-volovik,ref-chan,ref-bonneau1,ref-chaichian,
ref-victoria2,ref-battistel,ref-andrianov,ref-altschul1,ref-altschul2}.
Various schemes were suggested for identifying a single correct result, including some potential symmetry
arguments or attempts to characterize the theory nonperturbatively; however,
all proposed nonperturbative methodologies that could have led to nonzero values of
the coefficients of the induced Chern-Simons at odd orders in a power series expansion
utterly failed at even orders.

Another feature of the Chern-Simons term is its apparent instability, and quite a bit of research has
gone into understanding how this instability might or might not be tamable, depending on how precisely the
Chern-Simons modification to the theory is implemented. This makes the term one of the most fascinating
in the SME. The Chern-Simons term changes the propagation of electromagnetic radiation to make it
birefringent even in vacuum. Between the right- and left-circularly-polarized modes, one of them has its
phase speed increased, and the other mode's phase speed is decreased.
Experimentally, the distinctive Chern-Simons birefringence signature does not appear,
even for waves that have traversed cosmological distances~\cite{ref-carroll1,ref-kost21,ref-mewes5}, and
the lack of such birefringence has enabled some exceedingly tight bounds on
the magnitude of the physical Chern-Simons term. Moreover, the birefringence is closely tied to the
instability, since for sufficiently long wavelength modes, the dispersion relation may be so strongly
modified that $\omega^{2}<0$; and an imaginary frequency is normally associated with runaway, exponentially growing
solutions of the field equations.

This paper is organized as follows. In section~\ref{sec-LV-backgrounds}, we introduce the Lorentz-violating
Chern-Simons term and discuss some of its unusual properties. The potential instability of this theory is
one of its most notable features, and we discuss what is understood about how the instability might be
remedied in several different contexts, including the important case in which the Chern-Simons term exists
because the electromagnetic field is coupled to a slowly varying spin-0 field. In section~\ref{sec-solutns},
we present an iterative, order-by-order solution of the equations of motion for the gauge and cosmological pseudoscalar
fields, in the presence of a uniformly moving charge $e$. However, the iterative solution encounters some difficulties,
including singularities in the fields' calculated strengths beyond certain orders. Section~\ref{sec-nonpt} then shows
how a partial resummation of the power series solutions can cure some---but not all---of these singularities.
Finally, section~\ref{sec-concl} summarizes our conclusions about the interpretation of the paper's results.

\section{Lorentz Violation and Slowly Varying Backgrounds}
\label{sec-LV-backgrounds}

The form taken by the SME Lagrange density for the photon sector, with a CPT-violating Chern-Simons term
as the only Lorentz-violating addition, is
\begin{equation}
\label{eq-L}
{\cal L}=-\frac{1}{4}F^{\mu\nu}F_{\mu\nu}
+\frac{1}{2}k_{AF}^{\mu}\epsilon_{\mu\nu\rho\sigma}F^{\nu\rho}A^{\sigma}
-j^{\mu}A_{\mu},
\end{equation}
so that $k_{AF}$ represents a preferred axial vector background.
If spatial isotropy is unbroken in the preferred rest frame of the cosmological evolution, we may
write $k_{AF}^{\mu}=(k,\vec{0}\,)$, so that the $k_{AF}$ term is proportional simply to
$\vec{A}\cdot\vec{B}$ In this form, the potential gauge invariance issues of this theory are
fairly evident. The Lagrange density (\ref{eq-L})
containing $\vec{A}\cdot\vec{B}$ changes under a gauge transformation; however, the change is
a total derivative, so that the integrated action for the theory does actually remain gauge invariant.
This is enough to ensure that the equations of the motion---a modified version of Maxwell's equations---involve
only the electric and magnetic fields, not the unphysical potentials.

The photon dispersion relation with a purely timelike $k_{AF}$ looks deceptively simple,
$\omega_{\pm}^{2}=p(p\mp 2k)$. As already noted, the splitting of the modes' energies can actually
make the frequency imaginary for modes that are spatially varying only on very large scales (for which $p<|2k|$).
Another way to see evidence of the incipient instability is via the energy functional for the theory.
The energy-momentum tensor for the theory (for a general $k_{AF}$) is~\cite{ref-carroll1}
\begin{equation}
\Theta^{\mu\nu}= -F^{\mu\alpha}F^{\nu}\,_{\alpha}+\frac{1}{4}g^{\mu\nu}
F^{\alpha\beta}F_{\alpha\beta} -\frac{1}{2}k_{AF}^{\nu}\epsilon^{\mu\alpha\beta\gamma}
F_{\beta\gamma}A_{\alpha}.
\end{equation}
That this tensor is asymmetric is an indication of the Lorentz violation. However, the key property
of interest for characterizing the instability is the energy density,
\begin{equation}
\label{eq-Theta00}
\Theta^{00}=\frac{1}{2}\vec{E}^{2}+\frac{1}{2}\vec{B}^{2}-k\vec{A}\cdot\vec{B},
\end{equation}
(reverting again to a purely timelike $k_{AF}$). As with the Lagrange density (\ref{eq-L}), the energy
density is not a gauge-invariant quantity; thus it is not physically observable on its own. However,
once again, an integrated quantity (in this case the total energy, the integral
of $\Theta^{00}$ over all space) is actually gauge symmetric. The term that causes the difficulty is
again an $\vec{A}\cdot\vec{B}$ term, and this is also the term responsible for the instability.
For certain helicity modes in the Fourier expansion of $\vec{A}$,
making the mode amplitudes large can make the $\vec{A}\cdot\vec{B}$
term arbitrarily negative, and if the mode wavelength is sufficiently long (again, $p<|2k|$)
the negative energy of the Lorentz-violating term will win out over the usual $\vec{B}^{2}$
contribution to the energy.

One potential manifestation of the apparent instability could be vacuum Cerenkov radiation, which would
normally be expected in any theory in which charged particles can move faster than the phase speed of
light. Just as a tachyonic scalar field theory with Lagrange density
\begin{equation}
{\cal L}_{{\rm tachyon}}=\frac{1}{2}\partial^{\mu}\phi\partial_{\mu}\phi+\frac{\mu^{2}}{2}\phi^{2}
\end{equation}
appears to have a dispersion relation $\omega=\sqrt{p^{2}-\mu^{2}}$, which is not necessarily real, and
which we know actually signals that the free-field point $\phi=0$ is not a physical vacuum state, because
the energy may be made arbitrarily negative by increasing the amplitude of field modes with
$p<|\mu|$---so likewise the energy of the Chern-Simons theory may be made more and more negative by increasing
the amplitude of long-wavelength modes of $A^{\mu}$. These modes, which can be outpaced by a moving charge,
are somewhere we might expect to see vacuum Cerenkov radiation.

From the first introduction of the Lorentz-violating Chern-Simons term in Ref.~\cite{ref-carroll1}, there
has been interest in finding a way to evade the possible instability. The first proposed solution involved
calculating the radiation field of a current source using a Green's function that supports acausal propagation.
The solutions thus obtained obey the correct equations of motion. However, a charged particle will always
start to radiate before it actually begins to move. This is not
especially problematic for long wave trains oscillating at radio frequencies, but it does not provide a sensible
description for the excitation of modes with very long wavelengths, for which radiation can begin
arbitrarily far in advance of actual acceleration.

Since the first explication of the Chern-Simons theory, there have been a number of
other approaches to the basic problem of taming the instability in the theory. Often, vacuum Cerenkov radiation
provides a natural context for understanding these issues.
When the Lorentz-violating $k_{AF}$ coefficient is spacelike, the issue is not so fundamental, but the solution is
still quite illuminating~\cite{ref-lehnert1,ref-lehnert2,ref-kaufhold}. In this case, there exists a frame
in which $k_{AF}$ is purely spacelike, $k_{AF}^{\mu}=(0,\vec{k}_{AF})$, and in this frame, the energy density
(which is not a frame-invariant quantity) shows no instability, since the $\vec{A}\cdot\vec{B}$ term that may
be made arbitrarily negative is absent. On this basis alone, it may be expected that the theory should be free
of runaway Cerenkov radiation, and in fact, detailed calculations show that this is the case. A charge (even
one that is initially stationary in the laboratory frame) will radiate and accelerate until its rest frame is precisely
the frame in which $k_{AF}$ is purely spacelike---that is, the frame in which the theory is manifestly stable.

Of course, it is also possible to modify the theory with a timelike $k_{AF}$ to eliminate the instability.
Adding a Proca mass term $m_{\gamma}\geq k_{AF}^{0}$ makes the dispersion relation positive definite
and ensures that the potentially problematic $\vec{A}\cdot\vec{B}$ cannot win out over the positive
semidefinite mass term in the total energy~\cite{ref-cambiaso,ref-colladay3}.
This theory does support slow-moving electromagnetic modes, so
vacuum Cerenkov radiation is typically present when charges are moving, but there is no runaway radiation, since
the radiation has a definite threshold. If the velocity of a charge is below the threshold, there is no
vacuum Cerenkov radiation at all. As a result, an energetic charge will initially radiate, until its
speed and energy fall below the threshold, after which radiation ceases.
This is the same kind of behavior seen in theories with Lorentz-violating
but CPT-preserving minimal SME terms in the photon sector~\cite{ref-altschul9,ref-altschul12,ref-altschul19}. 

Perhaps the strangest case of ``stabilization'' of the Chern-Simons theory actually occurs without any
extra modifications to the Lagrange density with the purely timelike Chern-Simons term.
Direct calculations of the fields of a uniformly moving charge show that there is no net energy radiated, regardless
of the charge's speed~\cite{ref-schober1}. The condition of uniform motion transforms the potentially unstable,
exponentially growing modes of the field into modes with finite wavelengths but carrying negative energies.
The negative energies carried in the modes with $p<|2k|$ precisely cancel the positive energies carried by the
shorter-wavelength modes~\cite{ref-altschul21,ref-decosta1}.

This method of solving directly for the fields, assuming that they are in a steady-state configuration and
are thus following the charge as it moves, has fairly broad applicability. It could be used to explore the
modifications generated by a Chern-Simons term when there is Cerenkov radiation present due to other effects
(such as the presence of a material medium, or CPT-even Lorentz violation in the vacuum). The same method will
also be used in this paper, to find the fields generated by a moving charge in a somewhat more general theory.

That more general theory of interest here offers another way of apparently stabilizing the Chern-Simons form
of Lorentz violation. As already noted, the presence of a time-varying field associated with the slow expansion
of the universe naturally creates a preferred timelike direction. In Ref.~\cite{ref-kost20}, a string-motivated
cosmological model with two slowly-varying scalar and pseudoscalar fields $M$ and $N$ was introduced, with
an axion-like coupling of $N$ to the Abelian gauge field,
\begin{equation}
{\cal L}_{N}={\cal L}_{{\rm grav}}+\sqrt{g}\left(-\frac{1}{4\Lambda}MF^{\mu\nu}F_{\mu\nu}
-\frac{1}{4\Lambda}NF^{\mu\nu}\tilde{F}_{\mu\nu}
+\Lambda^{2}\frac{\partial^{\mu}A\partial_{\mu}A+\partial^{\mu}B\partial_{\mu}B}{4B^{2}}\right),
\end{equation}
where $\tilde{F}^{\mu\nu}=\frac{1}{2}\epsilon^{\mu\nu\rho\sigma}F_{\rho\sigma}$,
 $\Lambda$ is nominally the Planck mass, and in the weak field limit,
$M\approx B$ and $N\approx -A$. Slow cosmological variation in the field $M$ corresponds to a varying
gauge coupling $\alpha=\Lambda/4\pi M$. More interesting here is the variation of the pseudoscalar $N$. Assuming
that $N$ varies slowly enough that anything beyond its first derivative may be neglected,
the pure photon sector of this theory is equivalent to that of a theory
with an effective $k_{AF}^{\mu}=\frac{e^{2}}{2\Lambda}\partial^{\mu}N$

However, what makes this effective Chern-Simons theory particularly remarkable is that the energetic stability
problem, associated with the $-k\vec{A}\cdot\vec{B}$ term in the energy density, does not exist for this theory.
The reason for this is that the presence of $N$ introduces another dynamical field, which will carry
energy-momentum and contribute to $\Theta^{00}$. The contributions to $\Theta^{00}$ arising from the dynamics
of $M$ and $N$ are mostly well behaved, but they also include a
$+\frac{e^{2}}{2\Lambda}(\partial^{0}N)\vec{A}\cdot\vec{B}$ term,
which precisely cancels the problematic term in (\ref{eq-Theta00}). This means that including the dynamics
of this additional axion-like field may cure the stability problems of the Chern-Simons theory, and so it
is natural to try to understand in detail how this stabilization might occur, at the level of the field solutions.

\section{Iterative Solutions of the Field Equations}
\label{sec-solutns}

If particles can possess Lorentz-violating energy-momentum relations, it may be possible for
charged particles to move faster than the the phase speed of light. Since the Chern-Simons
term changes the dispersion relations for electromagnetic waves---in particular, slowing one
polarization mode down---vacuum Cerenkov radiation is a natural possibility in this theory.
However, there is an iterative algorithm for
determining the electric and magnetic fields of a moving point charge in the modified theory, and 
studies of the symmetry properties of this algorithm have shown that in the case of a
timelike Chern-Simons coefficient, there is zero radiation power loss from a uniformly moving
charge. The reason there is no energy loss is the cancellation between long- and short-wavelength modes
mentioned above.

We shall now generalize the iterative analysis, so that it will provide solutions to systems in which
a moving charge may generate not just electric and magnetic fields, but also excitations of the spin-0 field
$N$. The fields are those of a charge $e$ moving in the $z$-direction with velocity $\vec{v}$, passing through
the origin at time $t=0$. If the fields are in a steady state,
following along the movement of the charge (as they do for realistic
Cerenkov radiation in materials), then the field excitations can only depend on time through the
combination $\vec{x}-\vec{v}t$; their only time dependences come from the movement of the whole field profile
at velocity $\vec{v}$. This simplifies the field equations quite a bit, since any time derivative
$\partial W/\partial t$ of a field $W$ may be replaced with a spatial derivative $-v(\partial W/\partial z)$.

Of course, the cosmological background $N$ will have a different time dependence,
$N^{(1,0,-1)}=\frac{2\Lambda}{e^{2}}k(t-t_{0})$, which is approximately linearly varying on the time scale of interest.
The notation $N^{(1,0,-1)}$ indicates that this is the ${\cal O}(k^{1}v^{0}\Lambda^{1})$ term in an expansion
of $N$ in powers of $k$, $v$, and $\Lambda^{-1}$. In general, each field will be expanded according to the scheme
\begin{equation}
W=\sum_{i=0}^{\infty}\sum_{j=0}^{\infty}\sum_{n=-1}^{\infty}W^{(i,j,n)},
\end{equation}
where each term $W^{(i,j,n)}$ is proportional to $k^{i}v^{j}\Lambda^{-n}$. The field equations can then be solving
iteratively, with increasing $i+j+n$.

Except for $N^{(1,0,-1)}$ (which describes the background cosmological solution), each of the field terms
$\vec{E}^{(i,j,n)}$, $\vec{B}^{(i,j,n)}$, and $N^{(i,j,n\geq 0)}$ represents a field profile
of the type described above, moving
along with the charge. Each term has its time dependence entirely through $\vec{x}-\vec{v}t$, so we need not write this time
dependence explicitly; all the fields will be evaluated at the
specific time $t=0$ when the moving charge is located at the origin.

The equations of motion for the electromagnetic fields are Maxwell's equations, which may be modified due to the
presence of the pseudoscalar $N$.
The homogeneous equations of motion are not modified, since they merely indicate that the electric and
magnetic fields may be derived from scalar and vector potentials. Thus
\begin{eqnarray}
\label{eq-E-curl}
\vec{\nabla}\times\vec{E} & = & -\frac{\partial\vec{B}}{\partial t} \\
\vec{\nabla}\cdot\vec{B} & = & 0.
\end{eqnarray}
However, the sourced equations are modified,
\begin{eqnarray}
\label{eq-E-div}
\vec{\nabla}\cdot\vec{E} & = & \rho-\frac{e^{2}}{\Lambda}(\vec{\nabla}N)\cdot\vec{B} \\
\label{eq-B-curl}
\vec{\nabla}\times\vec{B} & = & \vec{J}+\frac{\partial\vec{E}}{\partial t}+
\frac{e^{2}}{\Lambda}\left(\frac{\partial N}{\partial t}
\vec{B}+\vec{\nabla}N\times\vec{E}\right),
\end{eqnarray}
so that they look like they possess an effective Chern-Simons term $k_{AF}^{\mu}=(k,\vec{0}\,)$, as well as additional
terms related to the dynamical part of $N$. [While $\vec{\nabla}N^{(1,0,-1)}=0$, the higher-order components $N^{(i,j,n)}$
will generally have nontrivial spatial dependences.]
In the purely-Chern-Simons-modified Ampere-Maxwell law,
\begin{equation}
\vec{\nabla}\times\vec{B}-\frac{\partial\vec{E}}{\partial t}=2k\vec{B}+\vec{J}
\end{equation}
the magnetic field becomes a source for itself,
behaving like an effective current source $\vec{J}_{{\rm eff}}=2k\vec{B}$. For comparatively
simple source configurations, the Maxwell's equations may be solved---sometimes
exactly~\cite{ref-schober2}, but more typically as a power series.

The equations (\ref{eq-E-curl}--\ref{eq-B-curl}) have to be supplemented with the equation of motion for $N$.
This would be derived from ${\cal L}_{N}$. When $A$ and $B$ are small,
$N=-A+{\cal O}(\Lambda^{-2})$ and $M=\Lambda/e^{2}=B+{\cal O}(\Lambda^{-2})$. Dropping all the terms
that do not involve $N$, any terms with suppression by higher powers of $\Lambda^{-1}$,
and the metric factors of $\sqrt{g}$, we are left with
\begin{equation}
\label{eq-LN}
{\cal L}\supset\frac{e^{4}}{4}(\partial^{\mu}N)(\partial_{\mu}N)-\frac{1}{4\Lambda}NF^{\mu\nu}\tilde{F}_{\mu\nu}.
\end{equation}
This gives an equation of motion for $N$,
\begin{equation}
-\frac{e^{4}}{2}\partial^{\mu}\partial_{\mu}N-\frac{1}{4\Lambda}F^{\mu\nu}\tilde{F}_{\mu\nu}=
-\frac{e^{4}}{2}\partial^{\mu}\partial_{\mu}N+\frac{1}{\Lambda}\vec{E}\cdot\vec{B}=0.
\end{equation}
Using the supposition that the time dependence of all the excitation fields comes purely from translation along
the $z$-direction, we can eliminate the temporal derivatives from $\partial^{\mu}\partial_{\mu}$,
\begin{equation}
\label{eq-box}
\partial^{\mu}\partial_{\mu}=v^{2}\frac{\partial^{2}}{\partial z^{2}}-\vec{\nabla}^{2}=
-\left[\frac{\partial^{2}}{\partial x^{2}}+\frac{\partial^{2}}{\partial y^{2}}+(1-v^{2})\frac{\partial^{2}}{\partial z^{2}}
\right],
\end{equation}
and if we can neglect terms of ${\cal O}(v^{3})$ and higher, we can approximate
$\partial^{\mu}\partial_{\mu}\approx-\vec{\nabla}^{2}$.

As just suggested, we shall henceforth only consider terms up to linear order in $v$; that means
field contributions $W^{(i,j,n)}$ with $j=0$ or $1$. If the
existence of independent dynamics for the $N$ field is responsible for stabilizing the theory, then that stabilization
mechanism should be evident for any $v>0$ (since vacuum Cerenkov radiation in this theory is not a
threshold phenomenon); and thus
the stabilization ought to be present even at the lowest nontrivial order in $v$. Note that this
means that the only electric field we shall need is the nonrelativistic Coulomb field, since any $k$-dependence
in $\vec{E}$ must arise via Faraday's law and a time-dependent $\vec{B}$---which
makes the $k$-dependent $\vec{E}$ terms automatically ${\cal O}(v^{2})$ or higher as well.

In Lorentz-invariant electrodynamics, the $\vec{E}$ and $\vec{B}$ of a moving charge are perpendicular. So the
lowest-order excitation for $N$ comes from
\begin{equation}
\label{eq-N-poisson}
\vec{\nabla}^{2}N^{(1,1,1)}=-\frac{2}{e^{4}\Lambda}\vec{E}^{(0,0,0)}\cdot\vec{B}^{(1,1,0)},
\end{equation}
with the unmodified $\vec{E}^{(0,0,0)}$ and the previously calculated~\cite{ref-altschul36}
\begin{equation}
\label{eq-B110}
\vec{B}^{(1,1,0)}=\frac{kev}{4\pi r}\left(2\cos\theta\,\hat{r}-\sin\theta\,\hat{\theta}\right).
\end{equation}
Solving the Poisson equation (\ref{eq-N-poisson}) gives
\begin{eqnarray}
N^{(1,1,1)} & = & \frac{1}{2\pi e^{4}\Lambda}
\int d^{3}x'\,\frac{\vec{E}^{(0,0,0)}(\vec{x}\,')\cdot\vec{B}^{(1,1,0)}(\vec{x}\,')}{|\vec{x}-\vec{x}\,'|} \\
& = & \frac{1}{2\pi e^{4}\Lambda}\int d^{3}x'\,\frac{\frac{ke^{2}v}{8\pi^{2}(r')^{3}}\cos\theta'}{|\vec{x}-\vec{x}\,'|}.
\end{eqnarray}
We may use the expansion of the Green's function in spherical harmonics
\begin{equation}
\frac{1}{|\vec{x}-\vec{x}\,'|}=\sum_{l=0}^{\infty}\sum_{m=-l}^{l}\frac{4\pi}{2l+1}\frac{r_{<}^{l}}{r_{>}^{l+1}}
Y_{lm}^{*}(\theta',\phi')Y_{lm}(\theta,\phi),
\end{equation}
so $N^{(1,1,1)}$ becomes
\begin{eqnarray}
N^{(1,1,1)} & = & \frac{kv}{16\pi^{3}e^{2}\Lambda}\int_{0}^{\infty}(r')^{2}\,dr'\,\frac{1}{(r')^{3}}
\sum_{l=0}^{\infty}\frac{r_{<}^{l}}{r_{>}^{l+1}}
\int_{0}^{\pi}\sin\theta'\,d\theta'\,\cos\theta' \\
& & \times \int_{0}^{2\pi}d\phi'\,\sum_{m=-l}^{l}\frac{4\pi}{2l+1}
Y_{lm}^{*}(\theta',\phi')Y_{lm}(\theta,\phi). \nonumber
\end{eqnarray}
Since $\cos\theta'\propto Y_{10}^{*}(\theta',\phi')$, only the $l=1$, $m=0$ term in the sum is nonzero,
and (substituting $u=\cos\theta'$)
\begin{eqnarray}
N^{(1,1,1)} & = & \frac{kv}{8\pi^{2}e^{2}\Lambda}
\int_{0}^{\infty}dr'\,\frac{r_{<}}{r'r_{>}^{2}}\int_{-1}^{1}du\,u^{2}\cos\theta \\
& = &\frac{kv}{12\pi^{2}e^{2}\Lambda}
\cos\theta\left[\int_{0}^{r}dr'\,\frac{1}{r^{2}}+\int_{r}^{\infty}dr'\,\frac{r}{(r')^{3}}
\right] \\
& = & \frac{kv}{8\pi^{2}e^{2}\Lambda r}\cos\theta.
\end{eqnarray}

The derivatives following from this are straightforward to calculate:
\begin{eqnarray}
\vec{\nabla}N^{(1,1,1)} & = & \frac{kv}{8\pi^{2}e^{2}\Lambda}
\left[\frac{\partial}{\partial r}\left(\frac{1}{r}\cos\theta\right)\hat{r}
+\frac{1}{r}\frac{\partial}{\partial\theta}\left(\frac{1}{r}\cos\theta\right)\hat{\theta}\right] \\
& = & \frac{kv}{8\pi^{2}e^{2}\Lambda}
\left(-\frac{1}{r^{2}}\cos\theta\,\hat{r}-\frac{1}{r^{2}}\sin\theta\,\hat{\theta}\right),
\end{eqnarray}
and
\begin{eqnarray}
\frac{\partial N^{(1,1,1)}}{\partial t}= -v\frac{\partial N^{(1,1,1)}}{\partial z} & = &
-\frac{kv^{2}}{8\pi^{2}e^{2}\Lambda}\frac{\partial}{\partial z}\left(\frac{z}{r^{2}}\right) \\
& = & -\frac{kv^{2}}{8\pi^{2}e^{2}\Lambda r^{2}}\left(1-2\cos^{2}\theta\right).
\end{eqnarray}
So that makes the source for the lowest term representing the back-reaction of $N$ onto $\vec{B}$
\begin{equation}
e^{2}\left(\frac{\partial N}{\partial t}\vec{B}+\vec{\nabla}N\times\vec{E}\right)\approx
e^{2}\left[-\frac{kv}{8\pi^{2}e^{2}\Lambda r^{2}}\left(\cos\theta\,\hat{r}+\sin\theta\,\hat{\theta}\right)\right]\times
\left(\frac{e}{4\pi r^{2}}\hat{r}\right).
\end{equation}
This consists solely of the $\vec{\nabla}N\times\vec{E}$ term.
The $\frac{\partial N}{\partial t}\vec{B}$ term is higher order by two powers of $v$, and we have already neglected
terms of that order coming from (\ref{eq-box}).

To see whether this might stabilize the state of the electromagnetic field, we need to get
the next iterated field $\vec{B}^{(1,1,2)}$, which  is determined by
\begin{equation}
\label{eq-B112}
\vec{\nabla}\times\vec{B}^{(1,1,2)}=\frac{kev}{32\pi^{3}\Lambda^{2}r^{4}}\sin\theta\,\hat{\phi}.
\end{equation}
This may be solved by the general method from Ref.~\cite{ref-schober1}. The radial dependence of $\vec{B}^{(1,1,2)}$
is constrained by dimensional analysis, so we set
\begin{equation}
\label{eq-Bnext-form}
\vec{B}^{(1,1,2)}=\frac{1}{r^{3}}\left[X(\theta)\hat{r}+Y(\theta)\hat{\theta}\right],
\end{equation}
so that
\begin{eqnarray}
\vec{\nabla}\times\vec{B}^{(1,1,2)} & = & \frac{1}{r^{4}}[-X'(\theta)-2Y(\theta)]\hat{\phi}
=\frac{kev}{32\pi^{3}\Lambda^{2}r^{4}}\sin\theta\,\hat{\phi} \\
\vec{\nabla}\cdot\vec{B}^{(1,1,2)} & = & \frac{1}{r^{4}}[-X(\theta)+\cot\theta\,Y(\theta)+Y'(\theta)]=0.
\end{eqnarray}
Solving for $Y(\theta)$ in the curl equation and inserting it into the divergence gives
\begin{equation}
\label{eq-X-ode}
X''(\theta)+\cot\theta\,X'(\theta)+2X(\theta)=-\frac{kev}{32\pi^{3}\Lambda^{2}}\cos\theta.
\end{equation}
The more general homogeneous differential equation
\begin{equation}
X''(\theta)+\cot\theta\,X'(\theta)+l(l-1)X(\theta)=0,
\end{equation}
[where $l=i-n=-1$ is determined by the $r^{-2+l}$
radial dependence of the toroidal function in (\ref{eq-Bnext-form})]
is a form of the Legendre equation, so it has solutions
\begin{equation}
X(\theta)=CP_{-l}(\cos\theta)+DQ_{-l}(\cos\theta),
\end{equation}
where the $P_{-l}(\xi)$ are the usual Legendre functions of the first kind (or Legendre polynomials when $l$ is an
integer) and the $Q_{-l}(\xi)$ are the Legendre functions of the second kind. The $Q_{-l}$ are usually not
part of physical solutions, because they all have logarithmic divergences at $\xi=\pm 1$---meaning on the $z$-axis for
$\xi=\cos\theta$. In this case, the required functions are
\begin{eqnarray}
P_{1}(\cos\theta) & = & \cos\theta \\
Q_{1}(\cos\theta)& = & \frac{1}{2}\log\left(\frac{1+\cos\theta}{1-\cos\theta}\right)\cos\theta-1.
\end{eqnarray}
For the inhomogeneous eq. (\ref{eq-X-ode}), there is also a particular solution
\begin{eqnarray}
X(\theta)=\frac{kev}{96\pi^{3}\Lambda^{2}}\log(\sin\theta)\cos\theta,
\end{eqnarray}
which is also singular for $\theta=0$ and $\pi$. By adding a $Q_{1}(\cos\theta)$ term with the right coefficient,
it is possible to eliminate the divergence on either the positive or negative $z$-axis, but not both. This leaves an
apparently unphysical solution.

The features of (\ref{eq-B112}) that make it unable to support a well-behaved solution can be identified more
clearly by solving the equation in a different way~\cite{ref-altschul36}. This method is applicable whenever
$\vec{\nabla}\times\vec{B}$ is a function of the form $f(r)\sin\theta\,\hat{\phi}$, because such a source
can be split into a collection of concentric spherical shells
of thickness $dR$, each carrying a surface current
$\vec{K}_{{\rm eff}}=dR\,[f(r)\sin\theta]\,\hat{\phi}$.
The magnetic field of such a current shell is well known (constant inside
and purely dipolar outside), so the contributions for the successive shells may be integrated to give the whole field,
provided that the solution exists and vanishes at spatial infinity. (If it does not, the integral will diverge.)
Applying this method to (\ref{eq-B112}) gives us
\begin{equation}
\label{eq-B112-div}
\vec{B}^{(1,1,2)}=\int_{0}^{r}dR\left(\frac{kev}{32\pi^{3}\Lambda^{2}R^{4}}\right)
\left[\frac{R^{3}}{3r^{3}}(3\cos\theta\,\hat{r}-\hat{z})
\right]+\int_{r}^{\infty}dR\left(\frac{kev}{32\pi^{3}\Lambda^{2}R^{4}}\right)\left(\frac{2}{3}\hat{z}\right),
\end{equation}
which diverges, because of the behavior of the integrand around $R=0$. For a more general power law
vortex source with $f(r)\propto r^{-\beta}$, it is clear from (\ref{eq-B112-div}) that a well-behaved
$\vec{B}$ will only exist for $1<\beta<4$. Since $\beta=4$ for the case of interest here, the desired field
profile does not quite exist; just as we already saw, there is a weak but unavoidable singularity on the $z$-axis.

In the case of a pure timelike Chern-Simons term, which is not associated with a separate
time-dependent field $N$, the remarkable cancellation that stabilizes the dynamics against Cerenkov losses
occurs between terms of different orders in $k$. We shall therefore also look at whether the problems we have
uncovered can be solved by including contributions from terms that are higher order in $k$. The next
term that could have an effect is ${\cal O}(k^{3}v)$. That there will be no effect from including
$\vec{B}^{(2,1,0)}$ can be seen in a couple different ways. Any magnetic field term with odd $j$ will
be purely azimuthal---thus perpendicular to $\vec{E}^{(0,0,0)}$---so it cannot serve directly as a
source for $N$. We also observe that the fundamental energetics of the theory can only depend on $|k|$, since
changing the sign of $k$ only switches the roles of the right- and left-circularly polarized modes; the switch
will lead to certain sign changes in the fields but should not affect total energy losses. Since the problematic term
in the energy includes one explicit factor of $k$, any nonzero contribution to the Cerenkov effect
at ${\cal O}(v)$ must come from a combination of fields proportional to an odd power of $k$.

We shall therefore look at how including $\vec{B}^{(3,1,0)}$ affects the structure of the fields as they
interact with $N$. The relationship between $\vec{B}^{(1,1,0)}$ and $\vec{B}^{(3,1,0)}$ is given by
\begin{equation}
\vec{\nabla}^{2}\vec{B}^{(i+2,1,0)}=-4k^{2}\vec{B}^{(i,1,0)}.
\end{equation}
To use the explicit integral solution of the Poisson equation,
\begin{eqnarray}
\vec{B}^{(3,1,0)} & = & \frac{k^{3}ev}{4\pi^{2}}\int dr'\,r'\!\int d\Omega'
\left[\cos\theta'\sin\theta'(\cos\phi'\hat{x}+\sin\phi'\hat{y})+(1+\cos^{2}\theta')\hat{z}\right] \\
& & \times\sum_{l=0}^{\infty}\frac{4\pi}{2l+1}\frac{r_{<}^l}{r_{>}^{l+1}}
\sum_{m=-l}^{l}Y^{*}_{l,m}(\theta',\phi')Y_{l,m}(\theta,\phi), \nonumber
\end{eqnarray}
requires regularization of the integral. So it is still easier to first calculate $\vec{B}^{(2,1,0)}$ (which may be
done by a straightforward pseudo-Amperean procedure~\cite{ref-schober1}),
\begin{equation}
\label{eq-B210}
\vec{B}^{(2,1,0)}=\frac{k^{2}ev}{2\pi}\sin\theta\,\hat{\phi}.
\end{equation}
From there, the same method that we used in (\ref{eq-Bnext-form}--\ref{eq-X-ode}) to calculate $\vec{B}^{(1,1,2)}$
can be followed again. The resulting solution is
\begin{equation}
\label{eq-B310}
\vec{B}^{(3,1,0)}=-\frac{k^{3}evr}{4\pi}\left(2\cos\theta\,\hat{r}-3\sin\theta\,\hat{\theta}\right),
\end{equation}
and this can then be used to calculate $N^{(3,1,1)}$, via
\begin{equation}
\label{eq-N311-poisson}
\vec{\nabla}^{2}N^{(3,1,1)}=\frac{k^{3}v}{4\pi^{2}e^{2}\Lambda r}\cos\theta,
\end{equation}
in which we have once again taken the dot product of the $\vec{B}^{(i,j,n)}$ with $\vec{E}^{(0,0,0)}$ to find
the right-hand-side.

Just as the magnetic field may sometimes be calculated as a superposition of the fields of spherical shells,
each carrying a perfectly dipolar surface current distribution, the Poisson equation (\ref{eq-N311-poisson})
may be solved by a method of superposing spherical shells carrying dipolar surface charge distributions. If
$\Phi_{R}\,dR$ is the electrostatic potential of a sphere carrying a surface charge $\sigma_{{\rm eff}}=-dR\,\cos\theta/R$,
\begin{equation}
\Phi_{R}(r)=\left\{
\begin{array}{ll}
-\frac{R^{2}}{3r^{2}}\cos\theta & r>R \\
-\frac{r}{3R}\cos\theta & r<R
\end{array}
\right.,
\end{equation}
then the solution of (\ref{eq-N311-poisson}) may be written
\begin{eqnarray}
N^{(3,1,1)} & = & \frac{k^{3}v}{4\pi^{2}e^{2}\Lambda}\int dR\,\Phi_{R} \\
\label{eq-N311}
& = & -\frac{k^{3}v}{4\pi^{2}e^{2}\Lambda}\left(\int_{0}^{r}dR\,\frac{R^{2}}{3r^{2}}\cos\theta+
\int_{r}^{\infty}dR\,\frac{r}{3R}\cos\theta\right),
\end{eqnarray}
and the second integral on the right-hand side of (\ref{eq-N311}) is logarithmically
divergent. Once again, we have arrived at an
infinite expression when $i+j+n$ has grown too large.

Of course, dimensional analysis suggests that $N^{(3,1,1)}$ ought to grow
approximately linearly with $r$ at large distances, so the divergence here is not necessarily a surprise.
In fact, by choosing an ansatz with additional logarithmic dependence on $r$, we find that
\begin{equation}
\label{eq-N311-log}
N^{(3,1,1)}=\frac{k^{3}v}{12\pi^{2}e^{2}\Lambda}r\log\left(\frac{r}{r_{0}}\right)\cos\theta
\end{equation}
is a solution of (\ref{eq-N311-poisson}). However, this is still somewhat unsatisfactory, since any positive
value of $r_{0}$ gives a solution. The reason is that
changing $r_{0}$ just adds a term proportional to $r\cos\theta$, which is a solution of the Laplace equation.
All this shows that an alternative approach is likely needed for dealing
with terms involving higher powers of $k$, and we shall introduce such an approach in section~\ref{sec-nonpt}.

However, interestingly, the symmetry arguments previously laid out in~\cite{ref-schober1} still ensure that
the electromagnetic energy that can escape to infinity still vanishes in this theory, even when the
the $N$-dependent sources for $\vec{E}$ and $\vec{B}$ are included. The inclusion of the additional field
does not change the key property that is responsible for the vanishing energy outflow---which is that 
the radial component
$\vec{S}\cdot\hat{r}$ of the modified Poynting vector, $S_{j}=\Theta^{j0}$, is always an odd function of
the $z$-coordinate.
(Note that, because of the Lorentz violating $k_{AF}$, $\vec{S}$ is not generally equal to the momentum density.
In fact, the asymmetry $\Theta^{\mu\nu}\neq\Theta^{\nu\mu}$ is a hallmark characteristic of Lorentz violation.)
$\vec{S}$ is not itself gauge invariant; to get a gauge-invariant expression describing the energy outflow, it
is necessary to take the integral of $\vec{S}\cdot\hat{r}$ over the sphere at $r\rightarrow\infty$. Because
the integrand is an odd function of $z$ (or equivalently, an odd function of $\cos\theta$), the integrated
quantity always vanishes.

Provided that the field profile for $N$ is azimuthally symmetric and an odd function of $\cos\theta$, the source
terms on the right-hand sides of (\ref{eq-E-div}) and (\ref{eq-B-curl}) will have the same symmetry structures
as they had in the absence of $N$. Conversely, when $\vec{E}$ and $\vec{B}$ have their expected symmetry properties,
$\vec{E}\cdot\vec{B}$ is independent of $\phi$ and odd in $\cos\theta$. Since $N$ is determined by solving
$\partial^{\mu}\partial_{\mu}N=(2/e^{4}\Lambda)\vec{E}\cdot\vec{B}$, the field $N$ has precisely the same symmetries
as $\vec{E}\cdot\vec{B}$ itself (even at higher order in $v$, when $\partial^{\mu}\partial_{\mu}$ cannot be
approximated by $-\vec{\nabla}^{2}$). The self-consistency of the reciprocal relations between $N$ and $F^{\mu\nu}$
thus ensure the symmetries of the electric and magnetic fields---and thus of $\vec{S}$---do not change even when
$k_{AF}$ is generated by the cosmological field $N$.

\section{Nonperturbative Solution for the Magnetic Field}
\label{sec-nonpt}

The forms of the usual nonrelativistic magnetic field $B^{(0,1,0)}$ and the higher-order
expressions (\ref{eq-B110}) and (\ref{eq-B210}--\ref{eq-B310}) suggest a common general form for
the magnetic field, in which each component of the field in spherical coordinates is a function of $r$ times a single
factor of $\cos\theta$ or $\sin\theta$. In fact, we can demonstrate that this is indeed the form taken by that
part of the field which is independent of $\Lambda$ and linear in $v$, but which encompasses all orders in $k$.

With an Ansatz of that form, the field is
\begin{equation}
\vec{B}^{({\rm all},1,0)}=\vec{B}^{({\rm e},1,0)}+\vec{B}^{({\rm o},1,0)},
\end{equation}
where $\vec{B}^{({\rm e},1,0)}$ and $\vec{B}^{({\rm o},1,0)}$ contain only even and odd powers of $r$, respectively.
They therefore take the Ansatz forms
\begin{eqnarray}
\vec{B}^{({\rm e},1,0)} & = & \sum_{i=0,\,i\,{\rm even}}^{\infty}a_{i}k^{i}r^{i-2}\sin\theta\,\hat{\phi} \\
\vec{B}^{({\rm o},1,0)}& = & \sum_{i=1,\,i\,{\rm odd}}^{\infty}k^{i}r^{i-2}\left(b_{i}\cos\theta\,\hat{r}
+c_{i}\sin\theta\,\hat{\theta}\right).
\end{eqnarray}
These must obey the modified Maxwell's equations for the magnetic field, order by order.

The divergences are simple.
$\vec{B}^{({\rm e},1,0)}$, as a purely azimuthal function with a magnitude independent of $\phi$,
is automatically divergenceless. In order to have $\vec{\nabla}\cdot\vec{B}^{({\rm o},1,0)}=0$, the
coefficients must obey the relation $c_{i}=-\frac{i}{2}b_{i}$.

The curl conditions are more intricate, however. Imposing $c_{i}=-\frac{i}{2}b_{i}$ and taking the curl
of $\vec{B}^{({\rm o},1,0)}$ gives
\begin{equation}
\vec{\nabla}\times\vec{B}^{({\rm o},1,0)}=\sum_{i}\left[-\frac{(i+1)(i-2)}{2}b_{i}\right]k^{i}r^{i-3}
\sin\theta\,\hat{\phi}.
\end{equation}
This is equal to $2k\vec{B}^{({\rm e},1,0)}$ if
\begin{equation}
b_{i}=-\frac{4}{(i+1)(i-2)}a_{i-1}.
\end{equation}
The curl of $\vec{B}^{({\rm e},1,0)}$,
\begin{equation}
\vec{\nabla}\times\vec{B}^{({\rm e},1,0)}=\sum_{i}k^{i}r^{i-3}\left[2a_{i}\cos\theta\,\hat{r}-(i-1)a_{i}\sin\theta\,
\hat{\theta}\right],
\end{equation}
is also needed. However, it is not quite the case that $\vec{\nabla}\times\vec{B}^{({\rm e},1,0)}
=2k\vec{B}^{({\rm o},1,0)}$, because the modified Ampere's law also involves the underlying current source itself.
This means that the initial coefficient $a_{0}$ is determined from the usual form of the
magnetic field of a moving charge, $a_{0}=\frac{ev}{4\pi}$. Then equating the coefficients of higher
powers of $k$ shows that $a_{i}=b_{i-1}$.

The recurrence relation for $a_{i}=a_{2p}$ is then
\begin{equation}
a_{2p}=\frac{-4}{2p(2p-3)}a_{2p-2}=\frac{(-1)^{p}}{p!\left(-\frac{1}{2}\right)_{p}}a_{0},
\end{equation}
where $(\xi)_{p}=\xi(\xi+1)(\xi+2)\cdots(\xi+p-1)$ is the Pochhammer symbol. This gives $\vec{B}^{({\rm e},1,0)}$
the form of a generalized hypergeometric function
\begin{equation}
\vec{B}^{({\rm e},1,0)}=\frac{ev}{4\pi r^{2}}\,_{0}F_{1}\left(;-\frac{1}{2};-k^{2}r^{2}\right)\sin\theta\,\hat{\phi}.
\end{equation}
By virtue of the hypergeometric function identities,
\begin{equation}
_{0}F_{1}\left(;\pm\frac{1}{2};-\xi^{2}\right) = \cos2\xi+\xi\sin2\xi\mp\xi\sin2\xi,
\end{equation}
we get the simple final form
\begin{equation}
\vec{B}^{({\rm e},1,0)}=\frac{ev}{4\pi r^{2}}\left[\cos(2kr)+2kr\sin(2kr)\right]\sin\theta\,\hat{\phi}.
\end{equation}
The sum giving $\vec{B}^{({\rm o},1,0)}$ can be evaluated similarly, although the result is slightly more elaborate,
\begin{equation}
\vec{B}^{({\rm o},1,0)}=\frac{kev}{2\pi r}\left\{\left[\frac{\cos(2kr)-1+2kr\sin(2kr)}{2k^{2}r^{2}}\right]
\left(\cos\theta\,\hat{r}+\frac{1}{2}\sin\theta\,\hat{\theta}\right)-\cos(2kr)\sin\theta\,\hat{\theta}\right\}.
\end{equation}

The next natural question is about the behavior of $N^{({\rm all},1,1)}$, sourced by
\begin{equation}
\vec{\nabla}^{2}N^{({\rm all},1,1)}=-\frac{2}{e^{4}\Lambda}\vec{E}^{(0,0,0)}\cdot\vec{B}^{({\rm all},1,0)}
=-\frac{kv}{8\pi^{2}e^{2}\Lambda r^{3}}\left[\frac{\cos(2kr)-1+2kr\sin(2kr)}{k^{2}r^{2}}\right]
\cos\theta.
\end{equation}
This may be solved by following the shell method introduced in the calculation of $N^{(3,1,1)}$.
Proceeding as in (\ref{eq-N311}), we have
\begin{equation}
N^{({\rm all},1,1)}=\frac{kv}{8\pi^{2}e^{2}\Lambda}\int_{0}^{\infty} dR\left\{
\begin{array}{ll}
\frac{R^{3}}{3r^{2}}\cos\theta & R<r \\
\frac{r}{3}\cos\theta & R>r
\end{array}
\right\}\left[\frac{\cos(2kR)-1+2kR\sin(2kR)}{k^{2}R^{5}}\right].
\end{equation}
Splitting the integral into its two regions, both constituent integrals can be performed,
although the results involve the sine and cosine integral functions,
\begin{eqnarray}
\Si(\xi) & = & \int_{0}^{\xi}d\eta\,\frac{\sin\eta}{\eta} \\
\Ci(\xi) & = & -\int_{\xi}^{\infty}d\eta\,\frac{\cos\eta}{\eta}.
\end{eqnarray}
In terms of these functions, the expression for $N^{({\rm all},1,1)}$ is
\begin{eqnarray}
N^{({\rm all},1,1)} & = & \frac{kv}{8\pi^{2}e^{2}\Lambda}\left[\frac{\cos(2kr)-1}{4k^{2}r^{3}}-\frac{5\cos(2kr)}{18r}
+\frac{5(k^{2}r^{2}-2)\sin(2kr)}{9kr^{2}}\right. \\
& & \left.-\frac{2\Si(2kr)}{3kr^{2}}-\frac{10kr^{2}\Ci(2kr)}{9}\right]\cos\theta. \nonumber
\end{eqnarray}
This is an improvement over the previous case, in which the unsatisfactory behavior of $N$ itself began with
$N^{(3,1,1)}$. The logarithmic behavior seen in (\ref{eq-N311-log}) is naturally included, through the $\Ci(2kr)$, but
as in Ref.~\cite{ref-schober2}, matching the solution found at leading order in $k$ to a nonperturbative
general solution transmutes the unknown scale factor $r_{0}$ from (\ref{eq-N311-log}) into a specific quantity
proportional to $k^{-1}$.
However, the nonperturbative solution still has leading-order $r^{-1}$ behavior, which means that the singularity in a
subsequent solution for $\vec{B}^{({\rm all},1,2)}$ will remain.

The nonperturbative resummation has thus eliminated one of the two divergences that bedeviled our earlier iterative
calculations. It is possible to obtain sensible expressions for the fields to all order is $k$. This suggests that
a similar resummation might work to address the divergence found at ${\cal O}(\Lambda^{-2})$. However, this turns out
not to be the case.

Although $k$ and $\Lambda^{-1}$ are both assumed to be small parameters, they have different units. If we try
to express the magnetic field as a sum to all orders in $\Lambda^{-1}$,
\begin{equation}
\vec{B}^{(1,1,{\rm all})}=\sum_{n=0}^{\infty}d_{n}\Lambda^{-n}r^{-n-1}\left(\cos\theta\,\hat{r}+\frac{n-1}{2}
\sin\theta\,\hat{\theta}\right),
\end{equation}
(where the coefficient of the $\hat{\theta}$ term has been selected to make each term divergenceless), problems will
arise, because of the the increasingly negative powers of $r$ that appear. In particular, the $n=2$ term is the
exterior field of a dipole, and so its curl is zero, except at $r=0$ where it has a strong singularity
(the derivative of a $\delta$-function). This means that the power series cannot extend past $n=2$, and so the
divergence in (\ref{eq-N311}) appears to be unavoidable.

\section{Conclusions}
\label{sec-concl}

Although the theory in which the Lorentz-violating Chern-Simons term arises as the derivative of a
cosmologically varying pseudoscalar field appears (based on inspection of its energy-momentum tensor) to
be better behaved than the pure Chern-Simons theory, we have encountered some puzzling results. Working to
successively higher powers of $k$ and $\Lambda^{-1}$ (those being the two very small parameters in the theory) we encountered
divergent expressions for terms in the magnetic field $\vec{B}$ and the spin-0 field $N$ that would
follow along with a charge in uniform motion. One of these divergences was evidently curable, since it was possible
to find resummed analytic formulas for $\vec{B}^{({\rm all},1,0)}$
and $N^{({\rm all},1,1)}$. However, the analogous divergence
encountered in the magnetic field at ${\cal O}(\Lambda^{-2})$ does not seem to have so simple a resolution.

The failure of the power series expansion at too high powers of $\Lambda^{-1}$ may actually not be too surprising.
The form (\ref{eq-LN}) for the Lagrangian governing the pseudoscalar field $N$ was only valid with sufficiently high
negative powers of $\Lambda$ neglected. At ${\cal O}(\Lambda^{-3})$, the equations governing the behavior of $N$ are
no longer universal and depend of the specifics of the underlying model. For the particular supergravity model
considered in Ref.~\cite{ref-kost20}, the higher-order equations are highly nonlinear and involve both $N$ and the
scalar $M$. It does not appear that the energetic stability of the theory should depend on these higher-order effects,
however; so the divergences we have uncovered may still be important to understanding the character of this theory.

If we take the solution for $\vec{B}^{(1,1,2)}$ from section~\ref{sec-solutns} that is regular on the positive $z$-axis,
\begin{equation}
\label{eq-B-defect}
\vec{B}^{(1,1,2)}=\frac{kev}{192\pi^{3}\Lambda^{2}r^{3}}\left\{2\left[\log(1+\cos\theta)\cos\theta-1\right]\hat{r}
-\left[\frac{3+2\cos\theta}{1+\cos\theta}-\log(1+\cos\theta)\right]\sin\theta\,\hat{\theta}\right\},
\end{equation}
we can envision a sort of interpretation for it.
As the particle propagates along the $z$-axis, it behaves almost like a zipper---metaphorically ``unzipping'' the
fields as it passes it and leaving behind a defect. The alternative solution that is regular on the negative
axis would correspond to ``advanced'' or time-reversed behavior, with the particle zipping up an existing singularity
in the fields as it moves along.

It is sometimes possible to have a field that diverges in a certain (measure zero) region without the situation
necessarily being pathological---for example, the weak divergence of the solution of the Dirac equation in the attractive
Coulomb potential at the origin.
However, this defect field (\ref{eq-B-defect}) clearly possesses a divergent energy, even with just the usual
$\frac{1}{2}\vec{B}^{2}$ taken into account. Yet this may actually be the key to understanding how the theory is stabilized. The
assumption underlying all our calculations is that there is a well-defined solution to the field equations in the
presence of a charged source that has been moving uniformly along the $z$-axis since $t=-\infty$. It may be that
such a steady solution simply does not exist. Regardless of its speed, a moving charge would be required to radiate
its energy away at a finite rate. The back-reaction due to the
charge's deceleration could then be responsible for smoothing
out the singular behavior of the field. This would be a new stabilization mechanism for this version of the
Chern-Simons theory, again unlike those that have been encountered in the other versions previously studied.

Alternatively,
in an even further modified electrodynamic theory, with some kind of short-distance regularization of the field
profiles, the difficulties with the singular
field strengths might be surmounted. This could mean, for example, using the nonlinear Born-Infeld
theory~\cite{ref-born-infeld} or the higher-derivative Bopp-Podolsky theory~\cite{ref-bopp,ref-podolsky}.
However, any further modifications to the theory to prevent the formation of overly strong fields would seemingly
render the modified theory incapable of answering our original questions about how the
version of the Chern-Simons theory with its Lorentz
violation derived from a slowly varying pseudoscalar was to be physically stabilized. Moreover,
while the specific example of the higher-derivative regularization in the Bopp-Podolsky theory (which incorporates 
a Pauli-Villars regulator directly into the photon field)
might resolve the divergent behavior up to some fixed
order in $\Lambda^{-1}$, other singularities would probably still occur at even higher orders.

We are thus left with a couple of plausible interpretations of our results. The peculiar singularities may be an
indication that the theory is inviable at a fairly fundamental level, in spite of its evident energetic stability.
Alternatively, they may be an indication only that uniform motion of charged particles over long stretches is not
possible in this theory. To evaluate whether this is a sensible interpretation will require looking at situations
involving actually
accelerating charges, which will probably entail significantly more elaborate calculations than have been performed up to
now in any version of the Lorentz-violating Chern-Simons theory.

\end{document}